# Designing Applications with Distributed Databases in a Hybrid Cloud

Evgeniy Pluzhnik[1], Oleg Lukyanchikov[2], Evgeny Nikulchev[1] & Simon Payain[1]

[1]*Moscow Technological Institute, Moscow, 119334, Russia,*
[2]*Moscow state university of instrument engineering and computer science, Moscow, 107996, Russia*

## Abstract

Designing applications for use in a hybrid cloud has many features. These include dynamic virtualization management and an unknown route switching customers. This makes it impossible to evaluate the query and hence the optimal distribution of data. In this paper, we formulate the main challenges of designing and simulation offer installation for processing.

## 1. Introduction

We present a set of technological approaches for efficient application development in a hybrid cloud infrastructure. Under the hybrid cloud infrastructure, we understand the distributed databases in the private and public cloud, as well as communication channels and programs providing the interaction between the components of the database.

A feature of the application is an intermediate layer that implements the connection of user requests to the location of distributed data. Presence of unknown destinations switching when using public cloud and mobile client makes it impossible to estimate the time of the algorithms. Here is why you want to use software technology to control all stages of the system. However, the hybrid infrastructure has many positive aspects of cloud computing: scalability and virtualization, and also due to the distribution of data to ensure the safety and security of data.

For the design of information systems in the cloud, there are the following problems:

- Impossible to assess the execution of individual queries and stream query;
- No general principles of designing systems with large amounts of data (BigData);
- Considerable amount educational and scientific data is semistructured (XML);

- No database migration technology to the cloud - to rewrite the code when moving to hybrid cloud;
- No commonly accepted principles virtualization management allocation of resources in the cloud
- The limitations associated with the use made obsolete protocol (TCP).

The task was to teach these features to develop the technology to create applications in a hybrid cloud.

Within the framework of these limitations, we have formulated the principles on which methods have been developed that provide a guaranteed quality of functioning of the application.

1. The design of the systems should be based on a preliminary study on the simulation and experimental models.

2. It is necessary to control the main parameters of the infrastructure.

3. The use of object-oriented technology modifications of database design.

4. Technology should provide the flexibility to change the structure of systems, data volume, number of requests.

## 2. Research and design of distributed systems

### 2.1 Experimental installation

An experimental test bench for simulation of hybrid clouds. He is a virtual farm with virtualization, the local server (for private clouds) and switching system (Fig. 1). When a client is randomly chosen route, the length of the delay and bandwidth. This allows you to simulate different access clients.

Virtualization environment is implemented VMware ESXi, installed using flash memory. Configured virtual switch Cisco Nexus 1000 and deployed 4 virtual machines on the physical disk server.

For organizations using cloud product family VMware vCloud, cloud computing allows you to organize at all levels. To create a cloud in the experimental stand on two servers SunFire hosts created VMware ESXi, established management system VCenter, installed VMware vCloud Director.

The presence of more than 15 physical Cisco switches and routers 29 Series 26 and Series 28, as well as the virtual switches Nexus, the functional network equipment, the use of dynamic routing protocols, technologies Vlan, trunk, QoS and other allows for a variety of network diagrams.

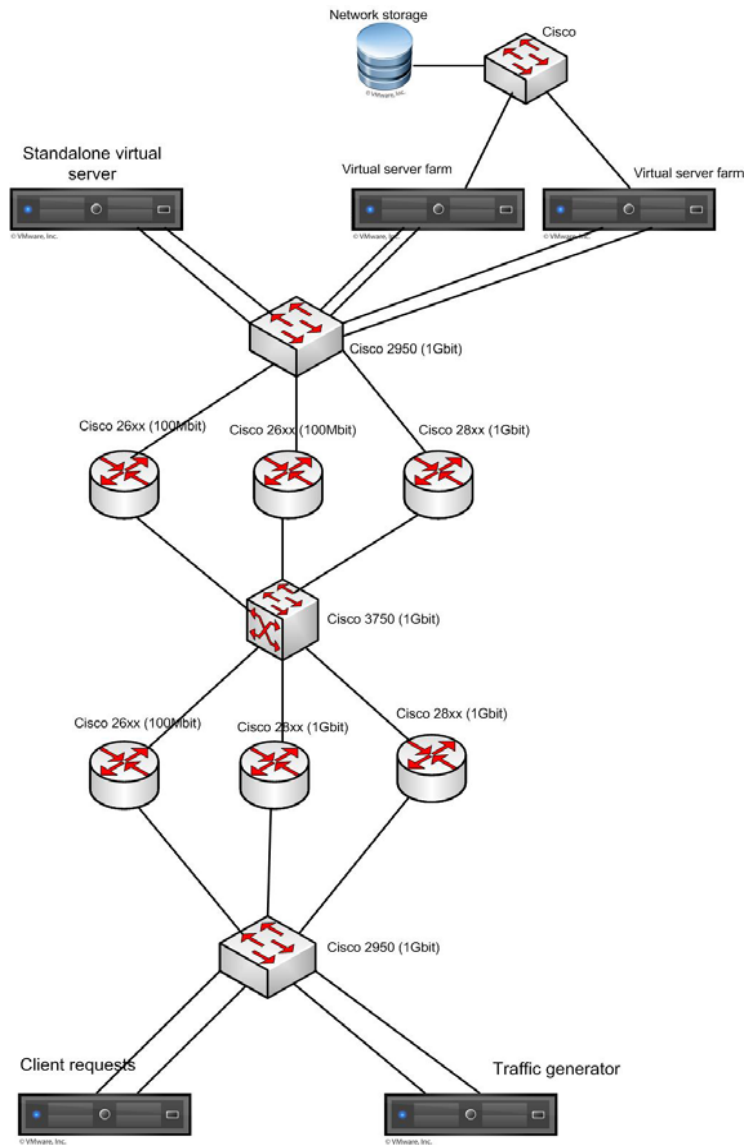
Fig.1: Experimental installation

## 2.2 Fast non-dominated sorting

Distributed data complicates the development of software, making it difficult and time-consuming to use common programming techniques. Despite the development of technologies such as .Net and Qt, developers have, eventually operate SQL queries and clearly prescribe the access to distributed data. In the context of widespread object-oriented development methodology and application systems at the same time a dominant position in the market relational DBMS attractive solution is the use of middleware software provides the necessary object-oriented interface to data stored under the control of a relational DBMS. To communicate with relational data objects with which developed software tools used programming technology object relational mapping (ORM). The essence of this technology is in accordance programming entity relational database object that is each field of a table is assigned a class attribute of the object.

The basic steps are the following:

1. Determination of the basic structure of the physical distribution of data in the hybrid cloud.

2. Development of the relational database structure.

3. Development of methods for data processing based on the physical location of the data.

4. Creating classes of objects, including data and methods for their treatment.

5. Modification of the structure as a result of an experimental study on the simulation bench.

6. Changing methods of processing inheritance

Depending on the task can be restructured to increase the speed of the most common queries, or make the data required to perform the most demanding requests in the public cloud.

Example research requests simulation setup is shown in Figure 2.

As a result of the emulation to stand for 30 hours were performed in 2194 random queries. The average duration of each query in seconds is shown in the figure.

According to the results it is possible to identify the most labor needs. Identified a series of simultaneous execution of queries that loads all. Those requests with less system load was performed without significant delays.

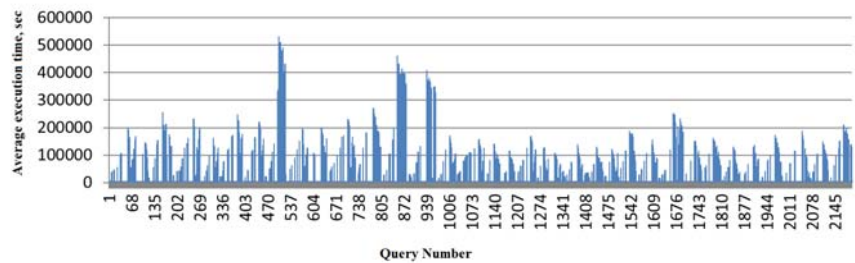

Fig.2: Experimental results

### 2.3 Control

The number of requests from different users can upload resources allocated for the task. For rapid expansion of computing power prompted for feedback on the state of the current query. In this case, to form a stream of complex queries can construct dynamic models in the form of a system of finite-difference equations. This makes it possible to assess the structural stability of the models in the test mode of operation.

In general, the introduction of feedback on the one hand, allows you to make corrective actions, on the other — occupies part of computational resources and communication channels, so it is necessary for optimal control problems where the quality criteria can serve to minimize the processing time or constrained resources (cloud technology service volume of maximum power determines the cost of the service). Note that the idea of feedback in computer systems has been elaborated [8], in particular in cloud technologies offer different ways of constructing a system of dynamic equations [9].

## 3. Conclusions

We formulate the basic principles of designing hybrid cloud systems in terms of processing large data. Proposed a kind of stand for experiments, experimental investigation and defines a set of technologies for the design. Ideas can be used for migration of existing systems in the cloud.

## Acknowledgements

The research work was supported by Russian Foundation for Basic Research under Grant No. 15-08-08935a.